\documentclass[aps,prl,amsfonts,amssymb,twocolumn,amsmath,preprintnumbers,nofootinbib,floatfix,showpacs,superscriptaddress]{revtex4-2}
\usepackage{graphicx}
\usepackage{epstopdf}
\usepackage{color}
\usepackage{hyperref}
\usepackage{bm}
\usepackage{printlen}
\usepackage{units}
\usepackage{bbm,pifont}
\usepackage{tikz}
\usepackage{amsmath}
\usepackage{amssymb}
\usepackage{amsthm}
\usepackage{commath}
\usepackage{amsfonts}
\usepackage[normalem]{ulem}
\usepackage[utf8]{inputenc}
\usepackage{epsfig}
\usepackage{latexsym}
\usepackage{xcolor}
\usepackage{subfigure}
\usepackage{array}
\usepackage{bbm}
\usepackage{hyperref,hypcap}
\usepackage{cancel}
\usepackage{ulem,dsfont}
\usepackage{braket}
\usepackage{algorithm}
\usepackage{algorithmic} 

\begin{document}
\title{Light Driven Spontaneous Phonon Chirality and Magnetization in Paramagnets}

\date{\today}

\author{Yafei Ren}
\affiliation{Department of Materials Science and Engineering, University of Washington, Seattle, Washington 98195, USA}
\affiliation{Department of Physics and Astronomy, University of Delaware, Newark, DE 19716, USA}

\author{Mark Rudner} 
\affiliation{Department of Physics, University of Washington, Seattle, Washington 98195, USA}

\author{Di Xiao}
\affiliation{Department of Materials Science and Engineering, University of Washington, Seattle, Washington 98195, USA}
\affiliation{Department of Physics, University of Washington, Seattle, Washington 98195, USA}

\begin{abstract}
Spin-phonon coupling enables the mutual manipulation of phonon and spin degrees of freedom in solids. 
In this study, we reveal the inherent nonlinearity within this coupling. 
Using a paramagnet as an illustration, we demonstrate the nonlinearity by unveiling spontaneous symmetry breaking under a periodic drive. 
The drive originates from linearly polarized light,  respecting a mirror reflection symmetry of the system. However, this symmetry is spontaneously broken in the steady state, manifested in the emergence of coherent chiral phonons accompanied by a nonzero magnetization. We establish an analytical self-consistent equation to find the parameter regime where spontaneous symmetry breaking occurs. Furthermore, we estimate realistic parameters and discuss potential materials that could exhibit this behavior. Our findings shed light on the exploration of nonlinear phenomena in magnetic materials and present possibilities for on-demand control of magnetization.
\end{abstract}

\maketitle

Spin-phonon coupling can significantly influence both the phononic and magnetic properties of materials~\cite{schaack1975magnetic, schaack1976observation, schaack1977magnetic, schaack1977magnetic1, thalmeier1977optical, ahrens1979phonon, capellmann1989microscopic, strohm2005phenomenological, inyushkin2007phonon, sheng2006theory, kagan2008anomalous, zhang2010topological, nova2017effective, juraschek2022giant, disa2020polarizing, afanasiev2021ultrafast, stupakiewicz2021ultrafast, mashkovich2021terahertz, luo2023large}. From the phonon perspective, this interaction can foster a strong coupling between phonons and external magnetic fields, leading to significant phonon energy splittings under a magnetic field~\cite{schaack1975magnetic, schaack1976observation, schaack1977magnetic, schaack1977magnetic1, thalmeier1977optical, ahrens1979phonon, capellmann1989microscopic}. It can also contribute to the phonon Hall effect~\cite{sheng2006theory, kagan2008anomalous, zhang2010topological}. 
From the spin perspective, phonons with nonzero angular momentum, commonly referred to as chiral phonons, can generate an effective magnetic field to influence spin dynamics~\cite{nova2017effective, juraschek2022giant, disa2020polarizing, afanasiev2021ultrafast, stupakiewicz2021ultrafast, mashkovich2021terahertz, luo2023large}. These chiral phonons can be excited, for instance, by circularly polarized light~\cite{juraschek2022giant, luo2023large}. 
As a result, spin-phonon coupling has recently attracted considerable attention due to their potential for light control of magnetization through lattice motion~\cite{nova2017effective, juraschek2022giant, disa2020polarizing, afanasiev2021ultrafast, stupakiewicz2021ultrafast, mashkovich2021terahertz, luo2023large}. 

In this Letter, we present a novel insight into the spin-phonon coupling by uncovering its inherent \textit{nonlinearity}. We reveal the emergence of a feedback loop resulting from their mutual interplay. Specifically, chiral phonons can induce spin magnetization, which, in turn, can stabilize the phonon chirality. This feedback loop can be activated by light that can induce large-amplitude and coherent lattice vibrations, thanks to recent advancements in THz laser pulses~\cite{disa2020polarizing, afanasiev2021ultrafast, stupakiewicz2021ultrafast, mashkovich2021terahertz} and nonlinear phononics~\cite{forst2011nonlinear, nicoletti2016nonlinear, mankowsky2016non, nova2017effective, subedi2021light, henstridge2022nonlocal}. 
Consequently, spin-phonon coupled systems offer an exceptional platform for exploring nonlinear dynamical phenomena~\cite{ohtsubo1999feedback, holden2014chaos, sciamanna2015physics, rudner2019self, aguirre2022recent} such as nonequilibrium spontaneous symmetry breaking~\cite{rudner2019self}, bifurcations~\cite{aguirre2022recent}, and chaos~\cite{sciamanna2015physics}. These phenomena hold significant practical importance as they enable the development of novel functionalities, allowing efficient control of spin using light on demand~\cite{nova2017effective, juraschek2022giant, disa2020polarizing, afanasiev2021ultrafast, stupakiewicz2021ultrafast, mashkovich2021terahertz, luo2023large}..

\begin{figure}
    \centering
    \includegraphics[width=8 cm]{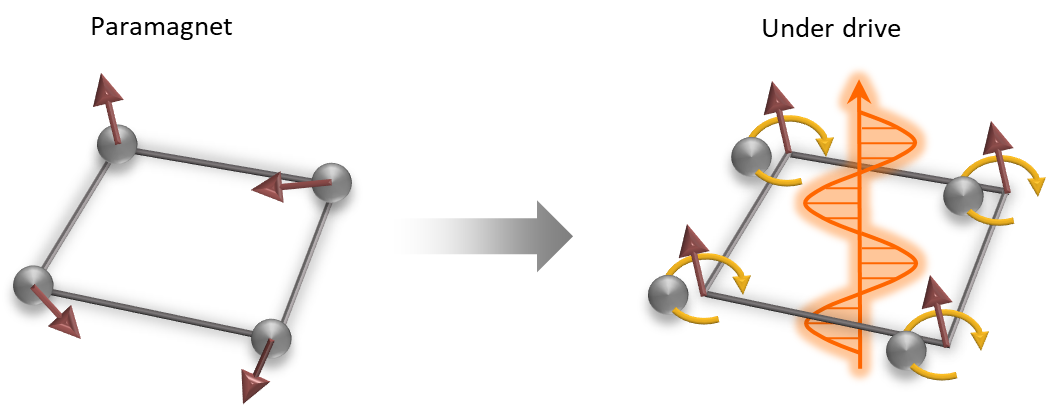}
    \caption{Illustration of spontaneous symmetry breaking. When a paramagnet with randomly oriented spins is exposed to a linearly polarized drive, despite maintaining mirror symmetry, the steady state exhibits a broken symmetry showing aligned spin and elliptically polarized lattice motion.}
    \label{Traj}
\end{figure}

Here, we demonstrate the nonlinearity by showing that  \textit{linearly} polarized light can induce chiral phonons and magnetization in an insulating paramagnet. 
This phenomenon exemplifies  nonequilibrium spontaneous symmetry breaking.
Specifically, we focus on a doubly degenerate infrared-active phonon mode that couples to linearly polarized light, which preserves a mirror reflection symmetry of the system.  
Intriguingly, we find that this symmetry can be spontaneously broken in the steady state, in which lattice motion becomes elliptically polarized, either left-handed or right-handed (randomly determined), displaying nonzero phonon angular momentum and spin magnetization (see Fig.~\ref{Traj}).
We subsequently elucidate the effects of various parameters, including the driving field strength, driving frequency, and damping, on symmetry breaking. Finally, we discuss potential material candidates and provide insights into the optimal conditions for experimental realization.

\textit{Spin-phonon coupling in a paramagnetic insulator.}---We decompose the three-dimensional atomic displacements into normal modes. 
Let us focus on doubly degenerate infrared-active phonon modes that naturally arise, for example, in systems with three or higher-fold rotation symmetry with the rotation axis denoted as the $z$ axis. The phonon modes carry electric dipoles $\bm{P}$ that are related to the phonon displacement $\bm{Q}=(Q_x, Q_y)$ by a tensor $\bm{Z}$, i.e., $P_i=Z_{ij} Q_j$, where the tensor element ${Z}_{ij}\equiv \partial P_i/\partial Q_j$ is related to the Born effective charge with $i(j)=x,y$.
We further assume that the system has a mirror symmetry, e.g., $\mathcal{M}_y$ about the $z$-$x$ plane, such that $\bm{Z}$
is diagonal in the coordinate system of $Q_{x,y}$. The dipoles for the $Q_{x}$ and $Q_{y}$ modes are along the $x$ and $y$ directions, respectively. 

When both modes are excited, the phonon can carry a nonzero angular momentum $\bm{l}= \bm{Q}\times \dot{\bm{Q}}$. The mirror and rotation symmetries guarantee that $\bm{l}$ can only be along the $\hat{z}$ direction and we thus only consider its $z$-component. 
Such a chiral lattice motion couples with local spins through the widely studied Raman-type spin-phonon coupling $\hat{\bm{S}} \cdot (\bm{Q}\times \dot{\bm{Q}})$~\cite{capellmann1989microscopic, sheng2006theory, kagan2008anomalous, zhang2010topological, komiyama2021universal, juraschek2022giant, footnote_note}, where $\hat{\bm{S}}$ represents the spin angular momentum operator. 
Through this coupling, a nonzero $l_z$ generates an effective Zeeman field for spins, which leads to a nonzero expectation value $\bm{S}=\langle \hat{\bm{S}} \rangle$. Reciprocally, a nonzero $\bm{S}$ generates an effective orbital magnetic field for phonons~\cite{capellmann1989microscopic, footnote_note1}, which stabilizes the phonon chirality.
Since the phonon angular momentum is along the $\hat{z}$ direction, only the $\hat{z}$ component of $\bm{S}$, denoted as $S\hat{z}$, can be nonzero. We thus omit the $\hat{x}, \hat{y}$ components of $\bm{S}$ in the following.

We describe the phonon dynamics in the presence of a linearly polarized light field using the Lagrangian~\cite{footnote_unit}
\begin{align}
\label{eq:Lagrangian}
    \mathcal{L}_{\rm ph}=\frac{1}{2} \dot{\bm{Q}}^2-\frac{\omega^2}{2}  \bm{Q}^2 - \frac{g S}{2} \hat{z} \cdot (\bm{Q}\times \dot{\bm{Q}}) + \bm{F}(t)\cdot \bm{Q},
\end{align}
where $\omega$ is the intrinsic oscillation frequency of the degenerate phonon modes and $g$ is the spin-phonon coupling constant.
The driving force is given by $F_j = E_i Z_{ij}$, where $\bm{E}=E(\cos \Omega t, 0, 0)$ is the electric component of the linearly polarized light field. 
The corresponding driving force is $\bm{F}(t)=F (\cos \Omega t, 0, 0)$, with the force amplitude $F=EZ_{xx}$ and driving frequency $\Omega=\frac{2\pi}{T_0}$, with $T_0$ being the period.
Note that the direction of $\bm{F}$ is along $x$, preserving the $\mathcal{M}_y$ symmetry of the system.
Supplementing the Euler-Lagrange equation associated with the Lagrangian in Eq.~(\ref{eq:Lagrangian}) with a phenomenological damping term, we obtain the equation of motion
\begin{align}
\label{EoM}
 \ddot{\bm{Q}} = - \omega^2 \bm{Q} + \bm{F} - g\dot{\bm{Q}}\times  S \hat{z} - \frac{g}{2} \bm{Q}\times \dot{S}\hat{z} - \dot{\bm{Q}}/\tau_p,
\end{align}
where $\tau_p$ is the damping coefficient related to the phonon lifetime. 
We assume $\tau_p$ is a constant parameter for simplicity. Through the equation of motion, Eq.~\eqref{EoM}, the net spin $S(t)$ influences the phonon dynamics analogously to an orbital magnetic field.

To model the spin dynamics, we focus on the regime of incoherent (relaxational) spin dynamics in which $S$ relaxes towards the equilibrium value $S_{\rm eq}$ with a phenomenological spin relaxation time $\tau_s$~\cite{crichton2019practical}:
\begin{align}
    \label{eq:SpinRelaxation}
    \dot{S}(t) = - [S(t) - S_{\rm eq}(l_z(t))]/\tau_s.
\end{align}
Here $S_{\rm eq}(l_z)=\tanh(\chi_p l_z)$ is the expectation value of spin in the presence of an effective Zeeman field $B_{\rm eff} = g l_z$ at temperature $T$ where $\chi_p = g/(2k_B T)$ is a material-specific parameter and $k_B$ is the Boltzmann constant (see Supplemental Materials for details).

\begin{figure}
    \centering
    \includegraphics[width=8 cm]{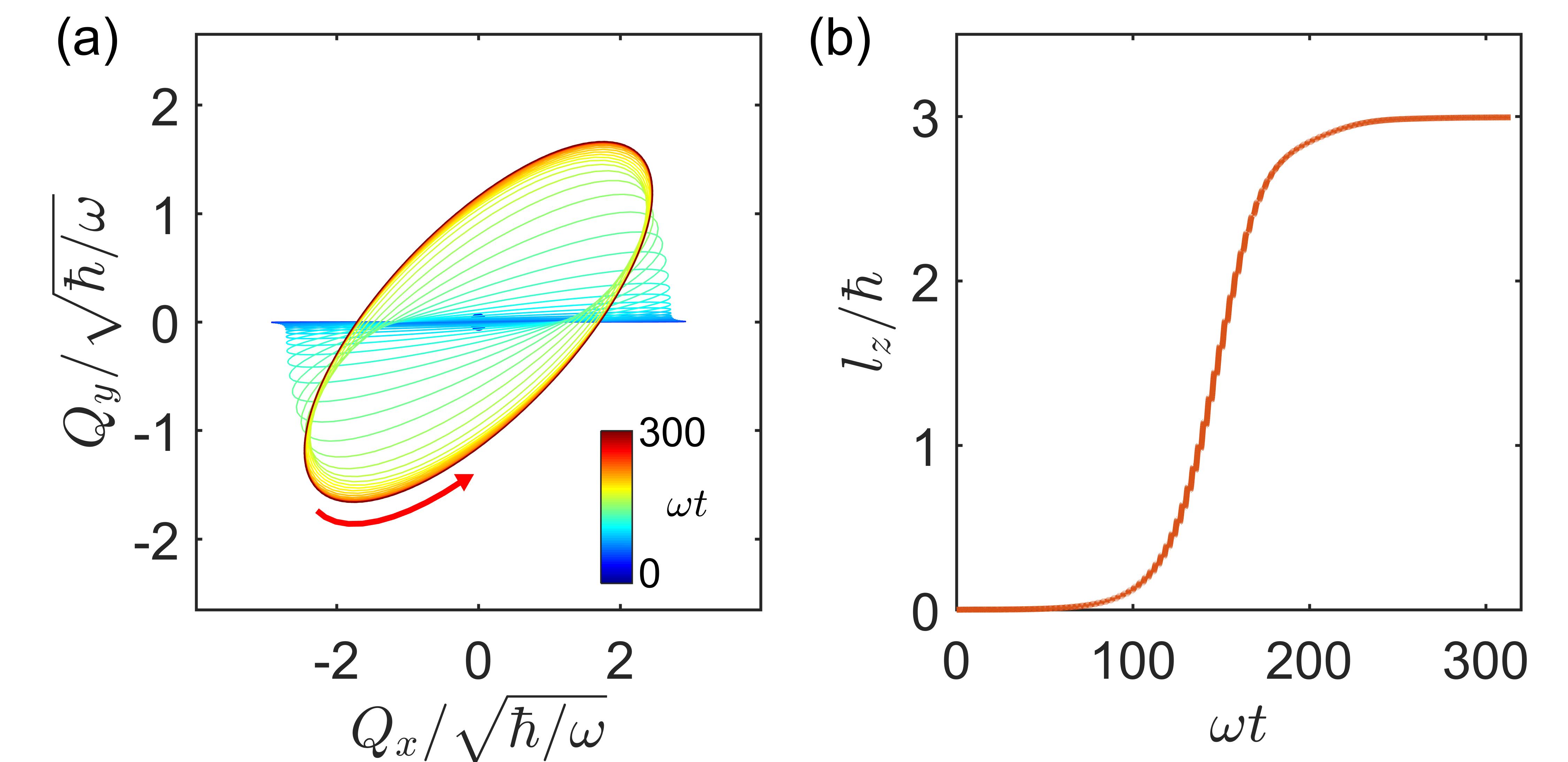}
    \caption{(a) Trajectory of $\bm{Q}(t)$ obtained by solving Eqs.~\eqref{EoM} and \eqref{eq:SpinRelaxation} simultaneously. The color gradation indicates time with blue representing the initial time. (b) Phonon angular momentum over time. In the calculation, $g=0.097 \omega$, $\chi_p=1.16\frac{1}{\hbar}$, $\omega \tau_p=10$, $F=0.4 \omega\sqrt{\hbar \omega}$, $\Omega=1.05\omega$, and $\omega \tau_s=10$.}
    \label{FigTraj}
\end{figure}

\textit{Spontaneous symmetry breaking.}---Equations~\eqref{EoM} and \eqref{eq:SpinRelaxation} are invariant under the mirror operation $\mathcal{M}_y$ about the $x$-$z$ plane.
By solving both equations simultaneously, the time evolution of $\bm{Q}$ and $S$ can be obtained. 
While it is natural to expect the steady state to be invariant under the mirror reflection, this symmetry can be {\it spontaneously} broken.
Figure~\ref{FigTraj}(a) illustrates an example where a trajectory $\bm{Q}(t)$ is plotted, with time $t$ represented by color gradation.
At the initial time $t=0$, the coordinate $\bm{Q}(0)$ is at the origin. 
To exhibit the instability, we initialize the system with a small kick along $Q_y$ leading to an extremely small $l_z$ at $t = 0$. 
As time evolves, $l_z$ starts growing as shown in Fig.~\ref{FigTraj}(b). Eventually, the trajectory spirals into a limit cycle forming a chiral steady state, which exhibits nonzero phonon angular momentum and spin magnetization.
This steady state is independent of the magnitude of the initial perturbative kick; we take $\dot{Q}_y(t=0)$ due to the initial kick to be $10^{-14}$ times the steady-state value of $|\dot{\bm{Q}}|$ in our simulation. 
Reversing the sign of $\dot{Q}_y(t=0)$ 
yields a steady state that is mirror-reflected with respect to $\mathcal{M}_y$ (hence with opposite angular momentum).
The chiral steady states break $\mathcal{M}_y$ symmetry spontaneously~\cite{Footnote_Fluctuation}, and are stable against small perturbations (see \textit{Stability} section below). 

\textit{Analytic self-consistent equation.}---Here we aim to find the phonon angular momentum in the steady state. 
Inspired by the example in Fig.~\ref{FigTraj}, we anticipate the steady state motion to be periodic with a period identical to that of the driving force.  We seek steady states with constant angular momentum, denoted as $\bar{l}_z$. 
Correspondingly, the spin $S$ is also a constant, denoted as $\bar{S}=S_{\rm eq}(\bar{l}_z)$.
Together with the assumptions above, for a fixed value of $\bar{l}_z$ the coupled equations of motion thus reduce to an effective driven {\it harmonic} (linear) oscillator; we seek self-consistent solutions such that the resulting steady-state value of angular momentum is equal to the supplied value of $\bar{l}_z$. 
For simplicity we initially assume the magnetization is small, $\bar{S}=S_{\rm eq}(\bar{l}_z)\approx \chi_p \bar{l}_z$.  
Later we will discuss the role of magnetization saturation.

We express the steady state of the effective linear oscillator as the real part of
\begin{align}
    \label{TrialSolution}
    (\bar{Q}_{x}, \bar{Q}_{y}) = (|\bar{Q}_{x}| e^{i\phi_x}, |\bar{Q}_{y}| e^{i\phi_y})e^{i\Omega t}, 
\end{align}
where $|\bar{Q}_{x,y}|$ and $\phi_{x,y}$ encode the amplitudes and phases of the $x$ and $y$ modes in the steady state.
This trial solution hosts a constant angular momentum $\bar{l}_z = \Omega {\rm Im}(\bar{Q}_{x} \bar{Q}_{y}^*)$ with $^*$ denoting complex conjugation.
By substituting Eq.~\eqref{TrialSolution} into Eq.~\eqref{EoM}, one can find that
\begin{align}
       (\bar{Q}_{x}, \bar{Q}_{y}) =& (\alpha, -i \beta \bar{l}_z) \frac{F}{\alpha^2 - (\beta\bar{l}_z)^2} e^{i\Omega t},
\end{align}
where $\alpha = \omega^2-\Omega^2 + i\Omega/\tau_p$, $\beta = -\chi \Omega$, and $\chi=g\chi_p$. We emphasize that $\bar{Q}_{x, y}$ depends on $\bar{l}_z$, and $\bar{l}_z$ is in turn a function of $\bar{Q}_{x, y}$.
We thus find a self-consistency equation for $\bar{l}_z$,
\begin{align}
\label{lzSC}
f(\bar{l}_z) \equiv \bar{l}_z |\alpha^2 -(\beta\bar{l}_z)^2|^2 + \Omega^2 F^2 {(\omega^2-\Omega^2) \chi} \bar{l}_z = 0.
\end{align}
The equation $f(\bar{l}_z)=0$ is a fifth-order polynomial equation, yielding five complex solutions in general. Among these solutions, only the real ones, referred to as ``fixed points,'' have physical significance. Note that $f(\bar{l}_z)$ and thus the fixed points are independent of the spin relaxation time $\tau_s$.

\begin{figure}
    \centering
    \includegraphics[width=8cm]{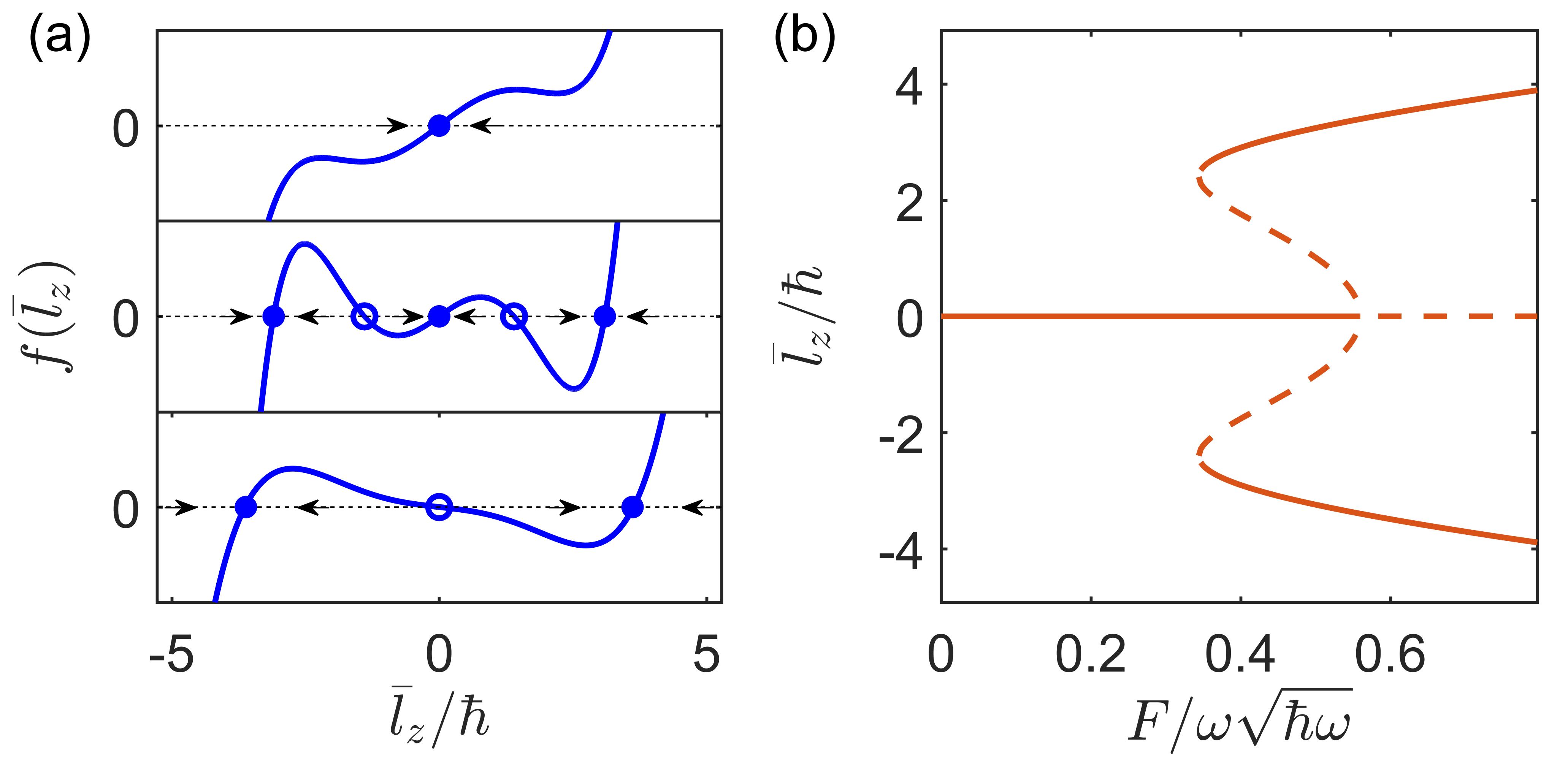}
    \caption{(a) $f(\bar{l}_z)$ vs $\bar{l}_z$ for various driving field strength $F$. From top to bottom, $F/\omega\sqrt{\hbar\omega}=0.13$, 0.45, and 0.93, respectively. Solid circles are stable fixed points whereas open circles are unstable. (b) Fixed points vs $F$. Solid and dashed lines are stable and unstable fixed points, respectively. 
    In the calculation, $\chi=0.11\frac{\omega}{\hbar}$, $\omega\tau_p=10$, $\omega \tau_s=100$, and $\Omega=1.156\omega$. }
    \label{FigPDOmegaF}
\end{figure}

By inspection of Eq.~(\ref{lzSC}), $\bar{l}_z=0$ is always a fixed point. 
This solution corresponds to a linearly polarized oscillation and respects the mirror symmetry $\mathcal{M}_y$ with vanishing magnetization. 
For a weak driving force $F$, $\bar{l}_z = 0$ is the only solution, see top panel of Fig.~\ref{FigPDOmegaF}(a) where $f(\bar{l}_z)$ is plotted. 
As the driving force increases, the number of fixed points increases to five, as shown in the middle panel. 
When the driving force increases further, the number of fixed points reduces to three, as shown in the bottom panel. 
The fixed points appear in pairs with opposite signs, as guaranteed by the mirror symmetry of the problem. 
The presence of fixed points with nonzero angular momentum, $\bar{l}_z\neq 0$, suggests that the system may exhibit spontaneous symmetry breaking; for confirmation, we next study the stability of the fixed points.
 
\textit{Stability analysis.}---The presence of spontaneous symmetry breaking requires the fixed point $\bar{l}_z=0$ to be unstable, i.e., an infinitesimal perturbation of ${l}_z$ can destroy it.  At the same time, there should be fixed points $\bar{l}_z\neq 0$ that are stable, i.e., robust against the perturbation.

To analyze the stability of the steady-state solutions $\bar{\bm{Q}}(t)$, we linearize the equation of motion around $\bar{\bm{Q}}(t)$ and $\bar{S}$ by introducing small perturbations $\bm{\eta}(t)$ and ${\zeta}(t)$, i.e., by setting $\bm{Q}(t)=\bar{\bm{Q}}(t) + \bm{\eta}(t)$ and ${S}(t)=\bar{{S}} + {\zeta}(t)$. By substituting $\bm{Q}(t)$ and $S(t)$ into Eqs.~\eqref{EoM} and \eqref{eq:SpinRelaxation} and retaining only up to linear terms in $\bm{\eta}$ and ${\zeta}$, we obtain the linearized equation of motion 
\begin{align}
    \frac{d}{dt} \Psi = H \Psi, 
\end{align}
where $\Psi=(\eta_x, \eta_y, \dot{\eta}_x, \dot{\eta}_y, \zeta)^{\rm T}$, and $H$ is a $5\times 5$ non-Hermitian matrix that has the same periodicity as the driving field, i.e., $H(t)=H(t+T_0)$.
This equation resembles a Floquet problem with a non-Hermitian Hamiltonian $H$. The stability of the solution $\bar{\bm{Q}}$ can be determined by calculating the logarithm of the eigenvalues of the time-evolution operator over a complete period, $\log \hat{\mathcal{T}}\exp(\int_0^{T_0} H dt)$, where $\hat{\mathcal{T}}$ indicates that the integral is time ordered.
If the real parts of the eigenvalues are all negative, a perturbation arising from a fluctuation of $\bm{\eta}$ and/or $\zeta$ will decay; the steady state is thus stable. Otherwise, the perturbation will increase exponentially, and the steady state is considered unstable.

By numerically evaluating and diagonalizing the Floquet operator, we identify that the fixed point with $\bar{l}_z=0$ in the top panel of Fig.~\ref{FigPDOmegaF}(a) is stable. Perturbations of the angular momentum decay to zero, as indicated by arrows. 
In the middle panel, three solutions are stable, as indicated by solid circles, whereas two solutions are unstable, as shown by open circles. In this case, even though there are symmetry-breaking stable steady states, fluctuations near $\bar{l}_z=0$ cannot induce spontaneous symmetry breaking as the $\bar{l}_z=0$ solution is also stable.
For even larger $F$ the $\bar{l}_z=0$ solution becomes unstable while the fixed points with $\bar{l}_z\neq 0$ are stable, as shown in the bottom panel.  In this case, spontaneous symmetry breaking appears.
In Fig.~\ref{FigPDOmegaF}(b) we show the fixed points as a function of the driving force, $F$.
The solid and dashed lines stand for stable and unstable fixed points, respectively.

\begin{figure}
    \centering
    \includegraphics[width=8 cm]{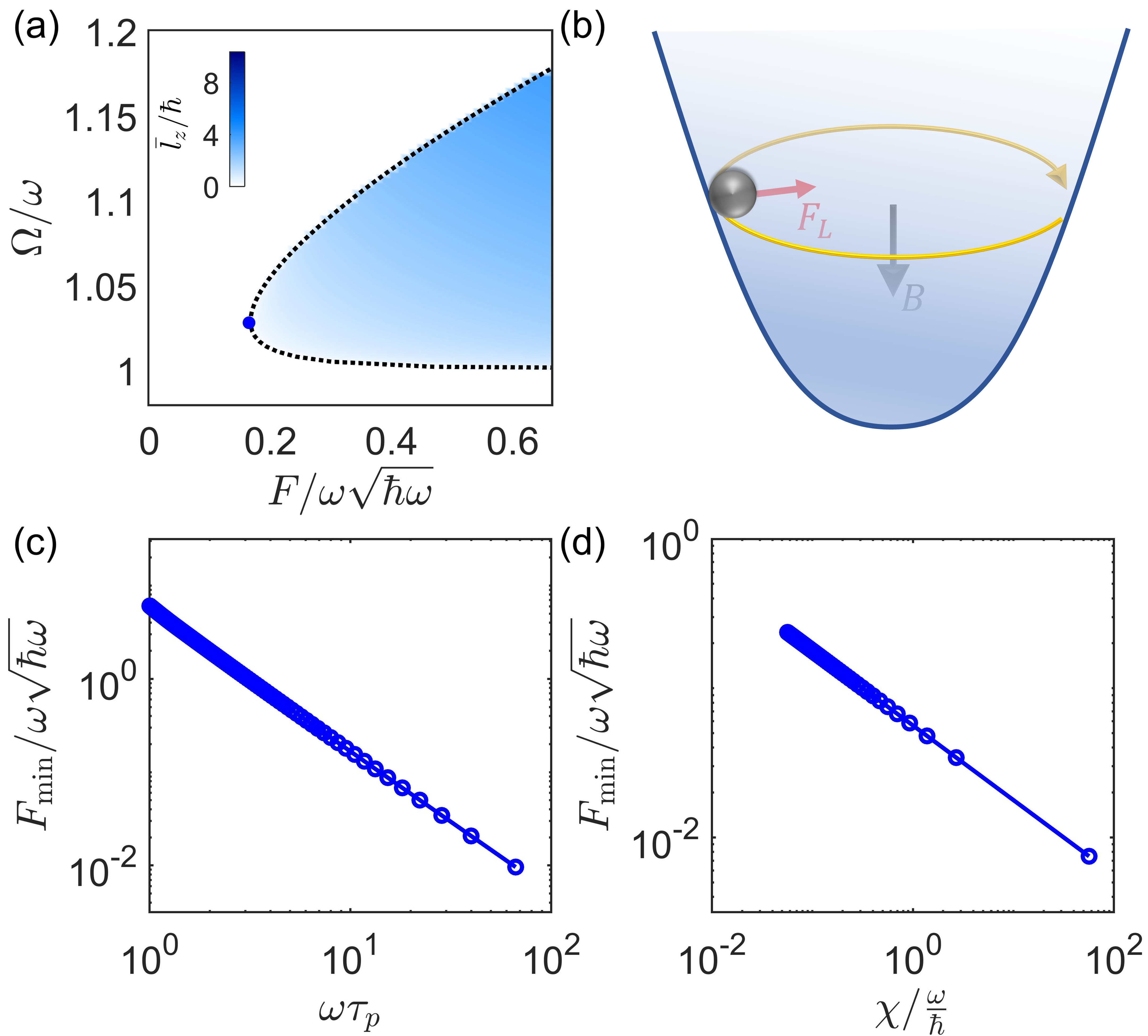}
    \caption{(a) Phase diagram in the parameter space spanned by driving force amplitude $F$ and driving frequency $\Omega$. 
    The dotted line shows the boundary between the phases with and without spontaneous symmetry breaking. The color corresponds to the magnitude of the phonon angular momentum in the steady state. The coordinate of the solid blue dot is $(F_{\rm min}, \Omega_{\rm min})$. In the calculation, $\chi=0.11\frac{\omega}{\hbar}$ and $\omega \tau_p=10$.
    (b) Illustration of the effective Lorentz force, $\bm{F}_{L}=-\dot{\bm{Q}}\times \bm{B}$ with $\bm{B}=gS\hat{z}\simeq \chi l_z \hat{z}$, from the spin-phonon coupling. Threshold of the field strength vs (c) $\tau_p$ and (d) $\chi$. }
    \label{FigThreshold}
\end{figure}

\textit{Phase diagram.---} The occurrence of spontaneous symmetry breaking also depends on the driving frequency $\Omega$. Figure~\ref{FigThreshold}(a) displays the phase diagram in the parameter space spanned by $F$ and $\Omega$. In the white region, there is no spontaneous symmetry breaking (i.e., throughout this region $\bar{l}_z=0$ is a stable fixed point). In the shaded region, spontaneous symmetry breaking appears. The color indicates the angular momentum $|\bar{l}_z|$ at the stable steady state obtained by solving the self-consistent equation. The boundary between the two phases is shown by the dashed line. 

The phase diagram reveals that the presence of spontaneous symmetry breaking necessitates $\Omega > \omega$.
An intuitive physical explanation is as follows.
As $\chi \propto g^2$ is positive, a clockwise rotation with $\bar{l}_z < 0$ generates an effective orbital magnetic field for phonons pointing downward, which induces a Lorentz force pointing inwards, parallel to the restoring force from the harmonic oscillator potential [Fig.~\ref{FigThreshold}(b)]. Consequently, the total restoring force increases, which induces a higher oscillation frequency. 
To resonantly excite this motion, a driving frequency $\Omega > \omega$ is required. The analysis for $\bar{l}_z >0$ is similar and one can find that the Lorentz force direction is independent of the sign of $\bar{l}_z$.

The phase diagram also shows that the presence of spontaneous symmetry breaking requires a finite field strength $F$. The smallest field strength for realizing it is shown by the blue dot in Fig.~\ref{FigThreshold}(a), the coordinate of which is $(F_{\rm min}, \Omega_{\rm min})$. The threshold value $F_{\rm min}$ is controlled by the damping parameter $1/\tau_p$, as depicted in Fig.~\ref{FigThreshold}(c) where $F_{\rm min}$ shows power-law dependence on the phonon lifetime $\tau_p$, approximately with $F_{\rm min} \propto \tau_p^{-3/2}$. Therefore, as $\tau_p$ becomes longer, a smaller field is required to realize the spontaneous symmetry breaking. Increasing $\chi$ can also reduce $F_{\rm min}$ as shown in Fig.~\ref{FigThreshold}(d), where $F_{\rm min} \propto \chi^{-1/2} \propto g^{-1}$, approximately.

\textit{Effects of spin saturation.}---The approximation $S_{\rm eq}(\bar{l}_z) \simeq \chi_p \bar{l}_z$ used to obtain the self-consistency equation for the steady states requires $|\chi_p \bar{l}_z| \ll 1$. This condition is satisfied near $(F_{\rm min}, \Omega_{\rm min})$ in Fig.~\ref{FigThreshold}(a), where $\bar{l}_z$ is small. Away from this regime, a full treatment using $S_{\rm eq}(l_z)=\tanh \chi_p l_z$ is necessary. 
In this case, we study the steady states numerically by solving Eqs.~\eqref{EoM} and \eqref{eq:SpinRelaxation} simultaneously. 
We find that the phase diagram in $\Omega$-$F$ space retains the same qualitative form as in Fig.~\ref{FigThreshold}(a), with quantitative agreement of the phase boundary in the region near $(F_{\rm min}, \Omega_{\rm min})$ [see Supplemental Materials].

\textit{Realistic material parameters.}---Various materials have been shown to exhibit strong spin-phonon coupling in both $f$-electron systems~\cite{schaack1976observation, schaack1977magnetic, schaack1977magnetic1, thalmeier1977optical, ahrens1979phonon, capellmann1989microscopic, strohm2005phenomenological, 4f1983LiTbF4}, such as CeCl$_3$, CeF$_3$, PrCl$_3$, and NdCl$_3$, and $d$-electron systems, e.g., CoTiO$_3$~\cite{chaudhary2023giant}. We take CeCl$_3$ as an example to estimate realistic parameters and calculate the driving-field amplitude and frequency for realizing such instability. A doubly degenerate infrared-active phonon mode in CeCl$_3$ has a frequency around $\omega = 2\pi\times$5 THz. The driving field frequency should be slightly above this value. 
We take $g$ as $2~$meV/$\hbar$ following relevant experiments and calculations~\cite{juraschek2022giant, schaack1976observation, schaack1977magnetic, thalmeier1977optical}. Given a temperature of $T \sim 10~$K, we find that $\chi \simeq 0.11 \frac{\omega}{\hbar}$ as used in our simulation. By controlling temperature, $\chi$ can change by orders since $\chi = g^2/2k_B T$.
The phonon lifetime can be estimated from the line width of the phonon spectrum, which is about $6\sim 40$ times smaller than the phonon energy; this corresponds to $\omega\tau_p$ in the order of $6 \sim 40$~\cite{luo2023large}. Decreasing the temperature, which can increase both $\chi$ and $\tau_p$, can reduce the threshold field strength. We find that $F_{\rm min}$ can decrease to the order of $0.02\omega\sqrt{\hbar\omega}$ at a low temperature of $1~$K with a phonon lifetime $\omega\tau_p\simeq 20$ from experiments~\cite{schaack1977magnetic}.
This threshold value is feasible in experiments.
By taking the Born effective charge as the elementary charge $e$ and the atomic mass as the nuclear mass of Cl, an external electric field ranging from $1\sim 10~$MV/cm can generate a driving force ranging from $0.03\sim 0.3 \omega\sqrt{\hbar\omega}$. 
Such spontaneous symmetry breaking can be experimentally measured since the chiral phonon-induced effective magnetic field is in the order of $10~$T when the angular momentum $\bar{l}_z$ reaches $2\hbar$, which can generate significant spin magnetization. 

\textit{Summary.}---In this work, we have used an insulating paramagnet to demonstrate the mechanism of spontaneous phononic chirality attributed to the inherent nonlinearity of spin-phonon coupling. 
Optimal material candidates should possess strong spin-phonon coupling manifested as large phonon energy splitting under a magnetic field and large magnetic susceptibility that typically can be found at a temperature slightly above the Curie or N\'eel temperature. 
Furthermore, a long phonon lifetime from a clean sample and weak phonon-phonon scattering can help to reduce the threshold of the driving field strength.

The spontaneous symmetry breaking leads to an upward or downward magnetization, introducing a binary degree of freedom. Fluctuations can induce either state randomly. Preference towards one of them can be introduced by deliberately disrupting the mirror symmetry by applying strain or in-plane electric fields, or by rotating the light polarization direction away from the mirror plane. It paves the way for engineering magnetic materials near the phase transition temperature by light and opens up promising avenues for potential applications.

The analytical study is supported by DOE Award No.~DE-SC0012509, and the numerical simulation is supported by
the Center on Programmable Quantum Materials, an Energy Frontier Research Center funded by DOE BES under award DE-SC0019443.
M.R. also acknowledges the Brown Investigator Award, a program of the Brown Science Foundation, the University of Washington College of Arts
and Sciences, and the Kenneth K. Young Memorial Professorship for support.


\begin{thebibliography}{37}
\bibitem[{foo({\natexlab{a}})}]{footnote_note}
	\bibinfo{note}{This coupling is referred to as spin-lattice coupling in some
		references. Here, we term it spin-phonon coupling, considering the fact that
		such coupling is valid only for the optical phonon modes of interest while
		the other phonon modes are neglected. This distinction is made to
		differentiate it from the coupling between spin and static lattice-structure
		deformation.}
\bibitem[{foo({\natexlab{b}})}]{footnote_note1}
	\bibinfo{note}{Here we consider the spatially averaged $\hat{\bm{S}}$, which
		acts as as a uniform orbital magnetic field for the phonons. Its fluctuations
		can induce phonon scattering that contributes to the damping of phonons,
		captured by the phenomenological damping term in Eq.~(\ref{EoM}).}
\bibitem[{Foo()}]{Footnote_Fluctuation}
	\bibinfo{note}{The initial nonzero velocity plays the role of a random
		fluctuation that determines the sign of the observed symmetry breaking.}
\end{thebibliography}

\end{document}